# A closed – form analytical solution to the radiation problem from a short dipole antenna above flat ground using spectral domain approach

S. Sautbekov*,
*Eurasian National University,5, Munaitpasov Str., Astana, Kazakhstan

sautbek@mail.ru

P. Frangos**, Ch. Christakis** and K. Ioannidi**
**Division of Information Transmission Systems and Materials Technology,
School of Electrical and Computer Engineering, National Technical University of Athens,
Iroon Polytechniou St.9, 157 73 Zografou, Athens, Greece
pfrangos@central.ntua.gr

*Abstract*—In this paper we consider the problem of radiation from a vertical short (Hertzian) dipole above flat ground with losses, which represents the well – known in the literature 'Sommerfeld radiation problem'. We end up with a closed – form analytical solution to the above problem for the received electric and magnetic field vectors above the ground in the far field area. The method of solution is formulated in the spectral domain, and by inverse three – dimensional Fourier transformation and subsequent application of the Stationary Phase Method (SPM) the final solutions in the physical space are derived. To our knowledge, the above closed – form solutions are *novel in the literature* for the Sommerfeld radiation problem. Finally, the physical interpretation for the received fields formulae derived this paper are provided.

*Index Terms*—Sommerfeld radiation problem, Spectral domain solution, Stationary Phase Method.

I.  INTRODUCTION

The so - called 'Sommerfeld radiation problem' is a well – known problem in the area of propagation of electromagnetic (EM) waves above flat lossy ground for obvious applications in the area of wireless telecommunications [1,2]. The classical Sommerfeld solution to this problem is provided in the physical space by using the so- called 'Hertz potentials' and it does not end – up with closed form analytical solutions. K. A. Norton [3] concentrated in subsequent years more in the engineering application of the above problem with obvious application to wireless telecommunications, and provided approximate solutions to the above problem, which are represented by rather long algebraic expressions for engineering use, in which the so – called 'attenuation coefficient' for the propagating surface wave plays an important role.

In this paper the authors take advantage of previous research work of them for the EM radiation problem in free space [4] by using the spectral domain approach.

Furthermore, in Ref. [5] the authors provided the fundamental formulation for the problem considered here, that is the solution in spectral domain for the radiation from a dipole moment at a specific angular frequency (ω) in isotropic media with a flat infinite interface. At that paper, the authors end – up with integral representations for the received electric and magnetic fields above or below the interface [Line of Site (LOS) plus reflected field – transmitted fields, respectively], where the integration takes place over the radial spectral coordinate $k_\rho$ . Then, in the present paper the authors concentrate to the solution of the classical 'Sommerfeld radiation problem' described above, where the radiation of a vertical dipole moment at angular frequency ω takes place above flat lossy ground [this is equivalent to the radiation of a vertical small (Hertzian) dipole above the flat lossy ground, as it will be explained by formula in the main text]. By using the Stationary Phase Method (SPM method, [6], [7]) integration over the radial spectral coordinate $k_\rho$ is performed and novel, to our knowledge, closed – form analytical solutions for the received electric and magnetic fields in the far field zone (where SPM method is applicable) are derived. Finally, physical interpretation of these novel closed – form analytical expressions are provided.

II. GEOMETRY OF THE RADIATION PROBLEM

The geometry of the problem is given in Fig. 1. Here a Hertzian (small) dipole with dipole moment $\underline{p}$ directed to positive x – axis, at altitude $x_0$ above the infinite, flat and lossy ground, radiates time – harmonic electromagnetic (EM) waves at angular frequency ω=2πf [exp(-iωt) time dependence is assumed in this paper]. Here the relative complex permittivity of the ground (medium 2) is $\varepsilon'_r = \varepsilon'/\varepsilon_0 = = \varepsilon_r + ix$ , where $x = \sigma/\omega\varepsilon_0 = 18 \times 10^9 \, \sigma/f$ , σ being the ground conductivity, f the frequency of radiation and $\varepsilon_0 = 8.854 \times 10^{-12}$ F/m is the absolute permittivity in vacuum or air. Then the wavenumbers of propagation of EM waves in air and lossy ground, respectively, are given by the following

$$k_{01} = \omega/c_1 = \omega\sqrt{\varepsilon_1\mu_1} = \omega\sqrt{\varepsilon_0\mu_0\varepsilon_{r1}\mu_{r1}} = \omega\sqrt{\varepsilon_0\mu_0} \quad (1)$$

$$k_{02} = \omega/c_2 = \omega\sqrt{\varepsilon_2\mu_2} = \omega\sqrt{\varepsilon_{r2}\mu_{r2}\varepsilon_0\mu_0} = k_{01}\sqrt{\varepsilon_r + ix} \quad (2)$$

The Maxwell equations for the time – harmonic EM fields considered above are given by

$$\begin{cases} \text{rot}\,\underline{E} - i\omega\mu_0\mu_r \underline{H} = 0 \\ \text{rot}\,\underline{H} + i\omega\varepsilon_0\varepsilon_r \underline{H} = \underline{j} \end{cases} \quad (3)$$

where $\underline{j}$ is current density (source of EM fields considered here).

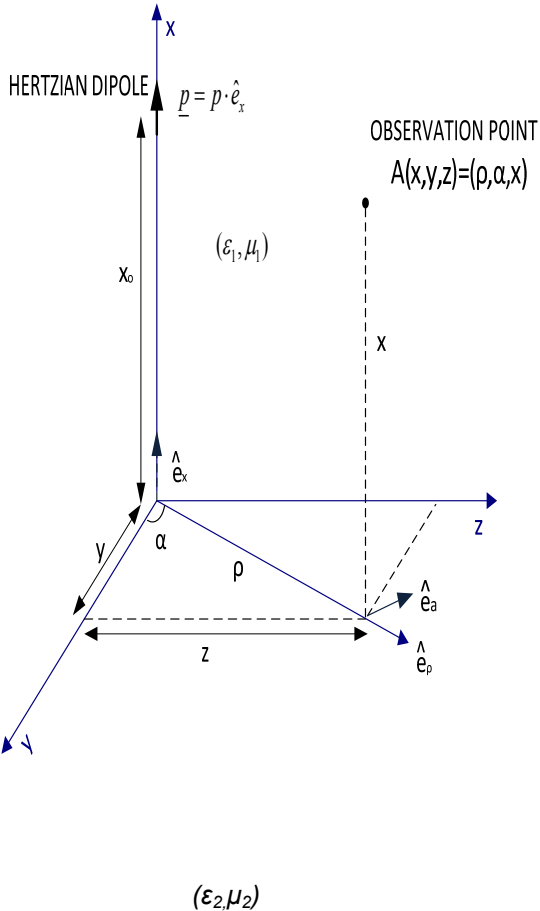

Fig. 1. Geometry of the radiation problem considered in this paper, namely radiation from a vertical Hertzian dipole at position $(x_0,0,0)$ above infinite, flat and lossy ground, as an interface situated at plane $x=0$ (Sommerfeld problem [1]-[3]).

III. FORMULATION OF THE SOMMERFELD RADIATION PROBLEM IN THE SPECTRAL DOMAIN :

INTEGRAL REPRESENTATION FOR THE RECEIVED ELECTRIC AND MAGNETIC FIELDS

### A. EM fields in terms of spectral domain current densities

Following [4]-[5], the EM field in physical space is derived from current density $\tilde{\underline{J}}$ in spectral domain and Green's function $\tilde{\psi}$, also in the spectral domain, through inverse three – dimensional (3D) Fourier transformation as following :

$$\underline{H} = -i\,\text{F}^{-1}\left[\tilde{\psi} \cdot (\underline{k} \times \tilde{\underline{J}})\right] \quad (4)$$

$$\underline{E} = -\frac{i}{\omega\varepsilon_r\varepsilon_0}\text{F}^{-1}\left\{\tilde{\psi}\left[\varepsilon_r\mu_r k_0^2 \tilde{\underline{J}} - <\underline{k},\tilde{\underline{J}}>\underline{k}\right]\right\} \quad (5)$$

where $\text{F}^{-1}$ is the inverse 3D Fourier Transform (FT) operator and

$$\tilde{\psi} = \left(k_{01}^2 - k^2\right)^{-1} = \left(k_{01}^2 - k_\rho^2 - k_x^2\right)^{-1} \quad (6)$$

is the 3D Green's function in spectral domain and cylindrical coordinates. Furthermore, by noting that, for the problem considered here, current density $\tilde{\underline{J}} = [\tilde{J}(k_\rho),0,0]$ has only $\rho$ – component, and that wavevector $\underline{k}=(k_\rho,k_\alpha=0,k_x)$ does not possess azimuthal $\alpha$ component, by performing the cross product and inverse FT operation of eq. (4) we obtain :

$$\underline{H}(\underline{r}) = -\frac{i}{(2\pi)^3}\hat{e}_\alpha \int_{k_\rho=0}^{\infty}\int_{\alpha=0}^{2\pi}\int_{k_x=-\infty}^{\infty} k_x \tilde{J}(k_\rho)\tilde{\psi}k_\rho \cdot \exp(i\underline{k}\cdot\underline{r})\cdot dk_\rho d\alpha dk_x \quad (7)$$

Similarly, by performing the inner product and inverse FT operation of eq. (5) we obtain :

$$\underline{E}(\underline{r}) = -\frac{i}{(2\pi)^3\varepsilon_r\varepsilon_0\omega}\int_0^\infty\int_0^{2\pi}\int_{-\infty}^{\infty}\left(\varepsilon_r\mu_r k_0^2 \hat{e}_\rho - k_\rho \underline{k}\right)\cdot \tilde{J}(k_\rho)\tilde{\psi}k_\rho \exp(i\underline{k}\cdot\underline{r})dk_\rho d\alpha dk_x \quad (8)$$

where

$$\underline{k} = (k_\rho, 0, k_x) = k_\rho \hat{e}_\rho + k_x \hat{e}_x \quad (9)$$

is the wavevector of propagation, and $\underline{r} = (\rho,\alpha,x)$ is the

point of observation (see Fig. 1), all in cylindrical coordinates. Furthermore, by taking into account eq. (9), eq. (8) for the received electric field can also be written in the form

$$\underline{E}(\underline{r}) = -\frac{i}{(2\pi)^3 \varepsilon_r \varepsilon_0 \omega} \int_0^\infty \int_0^{2\pi} \int_{-\infty}^\infty (\varepsilon_r \mu_r k_0^2 - k_\rho^2) \hat{e}_\rho -$$

$$-k_\rho k_x \hat{e}_x) \cdot \tilde{J}(k_\rho) \tilde{\psi} k_\rho \exp(i\underline{k} \cdot \underline{r}) dk_\rho d\alpha dk_x \quad (10)$$

Furthermore, in order to integrate expressions (7) and (10) with respect to azimuthal variable $\alpha$, we take into account that

$$\underline{k} \cdot \underline{r} = k_x x + k_\rho \rho \cdot \cos(\alpha - \beta) \quad (11)$$

Then, by using the following identities for Bessel functions

$$\frac{1}{2\pi} \int_0^{2\pi} \exp(ik_\rho \rho \cos \alpha) d\alpha = J_0(k_\rho \rho)$$

$$\int_0^\infty J_0(k_\rho \rho) dk_\rho = \frac{1}{2} \int_{-\infty}^\infty H_0^{(1)}(k_\rho \rho) dk_\rho \quad (12)$$

where $J_0$ is the Bessel function of first kind and zero order and $H_0^{(1)}$ is the Hankel function of first kind and zero order, we obtain

$$\underline{H}(\underline{r}) = -\frac{i}{8\pi^2} \hat{e}_\alpha \int_{k_\rho=-\infty}^\infty \int_{k_x=-\infty}^\infty k_x \tilde{J}(k_\rho) \tilde{\psi} k_\rho \cdot$$

$$\cdot H_0^{(1)}(k_\rho \rho) \exp(ik_x x) dk_\rho dk_x \quad (13)$$

$$\underline{E}(\underline{r}) = -\frac{i}{8\pi^2 \omega \varepsilon_r \varepsilon_0} \int_{k_\rho=-\infty}^\infty \int_{k_x=-\infty}^\infty ((\varepsilon_r \mu_r k_0^2 - k_\rho^2) \hat{e}_\rho -$$

$$-k_\rho k_x \hat{e}_x) \tilde{J}(k_\rho) \tilde{\psi} k_\rho H_0^{(1)}(k_\rho \rho) \exp(ik_x x) dk_\rho dk_x \quad (14)$$

B. *Formulation of the boundary value problem*

For the problem considered in this work (Fig. 1, above), we now use eqs. (13) and (14) above, to write the appropriate expressions for the reflected (R) and transmitted (T) EM field, as following :

$$\underline{H}^R(\underline{r}) = -\frac{i}{8\pi^2} \hat{e}_\alpha \int_{k_\rho=-\infty}^\infty \int_{k_x=-\infty}^\infty k_x \tilde{J}_1(k_\rho) \tilde{\psi}_1 k_\rho \cdot$$

$$\cdot H_0^{(1)}(k_\rho \rho) \exp(ik_x x) dk_\rho dk_x \quad (15)$$

$$\underline{E}^R(\underline{r}) = -\frac{i}{8\pi^2 \omega \varepsilon_{r1} \varepsilon_0} \int_{k_\rho=-\infty}^\infty \int_{k_x=-\infty}^\infty ((\varepsilon_{r1} \mu_{r1} k_{01}^2 - k_\rho^2) \hat{e}_\rho -$$

$$-k_\rho k_x \hat{e}_x) \tilde{J}_1(k_\rho) \tilde{\psi}_1 k_\rho H_0^{(1)}(k_\rho \rho) \exp(ik_x x) dk_\rho dk_x \quad (16)$$

$$\underline{H}^T(\underline{r}) = -\frac{i}{8\pi^2} \hat{e}_\alpha \int_{k_\rho=-\infty}^\infty \int_{k_x=-\infty}^\infty k_x \tilde{J}_2(k_\rho) \tilde{\psi}_2 k_\rho \cdot$$

$$\cdot H_0^{(1)}(k_\rho \rho) \exp(ik_x x) dk_\rho dk_x \quad (17)$$

$$\underline{E}^T(\underline{r}) = -\frac{i}{8\pi^2 \omega \varepsilon_{r2} \varepsilon_0} \int_{k_\rho=-\infty}^\infty \int_{k_x=-\infty}^\infty ((\varepsilon_{r2} \mu_{r2} k_{02}^2 - k_\rho^2) \hat{e}_\rho -$$

$$-k_\rho k_x \hat{e}_x) \tilde{J}_2(k_\rho) \tilde{\psi}_2 k_\rho H_0^{(1)}(k_\rho \rho) \exp(ik_x x) dk_\rho dk_x \quad (18)$$

where $k_{01}$ and $k_{02}$ are given by eqs. (1) and (2) above,

$$\tilde{\psi}_1 = \frac{1}{k_{01}^2 - k_\rho^2 - k_x^2} \quad (19)$$

$$\tilde{\psi}_2 = \frac{1}{k_{02}^2 - k_\rho^2 - k_x^2} \quad (20)$$

$\underline{\tilde{J}}_1 = [\tilde{J}_1(k_\rho), 0, 0]$, $\underline{\tilde{J}}_2 = [\tilde{J}_2(k_\rho), 0, 0]$ are the Fourier components of surface current density. Furthermore, the line-of-sight (LOS) EM field of the Hertzian dipole (in the *far field*, as it will be explained in Section III, below) is given by [6,7]

$$H_\alpha^{LOS}(r, \theta) = \frac{\omega^2 p}{4\pi} \sqrt{\varepsilon_0 \mu_0} \frac{\exp(ikr)}{r} \sin \theta \quad (21)$$

where spherical coordinates (r,θ) are given in terms of cylindrical coordinates (ρ,x) [see Fig. 1, above] by

$$r \approx \rho + \frac{(x-x_0)^2}{2\rho} \quad (22)$$

$$\theta = \frac{\pi}{2} - \tan^{-1}(\frac{x}{\rho}) \quad (23)$$

and

$$\underline{E}^{LOS}(r,\theta) = \zeta H_\alpha^{LOS} \cos\theta \, \hat{e}_\rho - \zeta H_\alpha^{LOS} \sin\theta \, \hat{e}_x \quad (24)$$

where $H_\alpha^{LOS}$ is given by eqs. (21) - (23) above.

Then, the total EM field in the regions x>0 and x<0 (see Fig. 1) is given by

$$\underline{H}(\underline{r}) = \begin{cases} \underline{H}^{LOS}(\underline{r}) + \underline{H}^R(\underline{r}), & x > 0 \\ \underline{H}^T(\underline{r}), & x < 0 \end{cases} \quad (25)$$

$$\underline{E}(\underline{r}) = \begin{cases} \underline{E}^{LOS}(\underline{r}) + \underline{E}^R(\underline{r}), & x > 0 \\ \underline{E}^T(\underline{r}), & x < 0 \end{cases} \quad (26)$$

Furthermore, by performing the integrations of expressions (15) – (18) over $k_x$, by using the *residue theory* [11], we obtain the following integral expressions for the EM fields :

In the upper half space (x>0) :

$$\underline{H}(\underline{r}) = \underline{H}^{LOS}(\underline{r}) - \frac{\hat{e}_\alpha}{8\pi} \int_{-\infty}^{\infty} k_\rho \tilde{J}_1(k_\rho) \, \mathrm{H}_0^{(1)}(k_\rho\rho) e^{i\kappa_1 x} dk_\rho \quad (27)$$

$$\underline{E}(\underline{r}) = \underline{E}^{LOS}(\underline{r}) - \frac{1}{8\pi\omega\varepsilon_{r1}\varepsilon_0} \hat{e}_\rho \int_{-\infty}^{\infty} \kappa_1 k_\rho \tilde{J}_1(k_\rho) \cdot \mathrm{H}_0^{(1)}(k_\rho\rho) e^{ik_1 x} dk_\rho +$$

$$+ \frac{1}{8\pi\omega\varepsilon_{r1}\varepsilon_0} \hat{e}_x \int_{-\infty}^{\infty} k_\rho^2 \tilde{J}_1(k_\rho) \, \mathrm{H}_0^{(1)}(k_\rho\rho) e^{i\kappa_1 x} dk_\rho \quad (28)$$

while for the lower half space (x<0) :

$$\underline{H}^T(\underline{r}) = \frac{\hat{e}_\alpha}{8\pi} \int_{-\infty}^{\infty} k_\rho \tilde{J}_2(k_\rho) \, \mathrm{H}_0^{(1)}(k_\rho\rho) e^{-i\kappa_2 x} dk_\rho \quad (29)$$

$$\underline{E}^T(\underline{r}) = -\frac{1}{8\pi\omega\varepsilon_{r2}\varepsilon_0} \hat{e}_\rho \int_{-\infty}^{\infty} \kappa_2 k_\rho \tilde{J}_2(k_\rho) \cdot \mathrm{H}_0^{(1)}(k_\rho\rho) e^{-ik_2 x} dk_\rho +$$

$$+ \frac{1}{8\pi\omega\varepsilon_{r2}\varepsilon_0} \hat{e}_x \int_{-\infty}^{\infty} k_\rho^2 \tilde{J}_2(k_\rho) \, \mathrm{H}_0^{(1)}(k_\rho\rho) e^{-i\kappa_2 x} dk_\rho \quad (30)$$

where

$$\kappa_1 = \sqrt{k_{01}^2 - k_\rho^2} \quad (31)$$

$$\kappa_2 = \sqrt{k_{02}^2 - k_\rho^2} \quad (32)$$

*C. Application of the Boundary Conditions (BC) - Solution for the uknown current densities at the interface in spectral domain*

We now apply the BC that at the interface (x=0) the tangential components of electric field $\underline{E}$ and magnetic field $\underline{H}$ must be continuous, namely

$$H_\alpha^{LOS} + H_\alpha^R = H_\alpha^T \quad (33)$$

$$E_\rho^{LOS} + E_\rho^R = E_\rho^T \quad (34)$$

where

$$H_\alpha^{LOS} = -\frac{1}{8\pi} \int_{-\infty}^{\infty} \frac{i\omega p k_\rho^2 e^{i\kappa_1 x_0}}{\kappa_1} \mathrm{H}_0^{(1)}(k_\rho\rho) dk_\rho \quad (35)$$

$$E_\rho^{LOS} = \frac{1}{8\pi\varepsilon_{r1}\varepsilon_0} \int_{-\infty}^{\infty} i\omega p k_\rho^2 e^{i\kappa_1 x_0} \mathrm{H}_0^{(1)}(k_\rho\rho) dk_\rho \quad (36)$$

$$H^R = -\frac{1}{8\pi} \int_{-\infty}^{\infty} k_\rho \tilde{J}_1(k_\rho) \, \mathrm{H}_0^{(1)}(k_\rho\rho) dk_\rho \quad (37)$$

$$E_\rho^R = -\frac{1}{8\pi\omega\varepsilon_{r1}\varepsilon_0} \int_{-\infty}^{\infty} \kappa_1 k_\rho \tilde{J}_1(k_\rho) \, \mathrm{H}_0^{(1)}(k_\rho\rho) dk_\rho \quad (38)$$

$$H_\alpha^T = \frac{1}{8\pi} \int_{-\infty}^{\infty} k_\rho \tilde{J}_2(k_\rho) \, \mathrm{H}_0^{(1)}(k_\rho\rho) dk_\rho \quad (39)$$

$$\underline{E}^T(\underline{r}) = -\frac{1}{8\pi\omega\varepsilon_{r2}\varepsilon_0} \int_{-\infty}^{\infty} \kappa_2 k_\rho \tilde{J}_2(k_\rho) \, \mathrm{H}_0^{(1)}(k_\rho\rho) dk_\rho \quad (40)$$

Then from eqs. (33) and (34) we find :

$$\frac{1}{8\pi}\int_{-\infty}^{\infty}\left(\frac{i\omega p k_\rho e^{i\kappa_1 x_0}}{\kappa_1}+\tilde{J}_1(k_\rho)\right)H_0^{(1)}(k_\rho\rho)k_\rho dk_\rho =$$
$$= -\frac{1}{8\pi}\int_{-\infty}^{\infty}\tilde{J}_2(k_\rho)H_0^{(1)}(k_\rho\rho)k_\rho dk_\rho$$
(41)

$$\frac{1}{8\pi\varepsilon_{r1}\varepsilon_0}\int_{-\infty}^{\infty}\left(-i\omega p k_\rho e^{i\kappa_1 x_0}+\tilde{J}_1(k_\rho)\kappa_1\right)\cdot H_0^{(1)}(k_\rho\rho)k_\rho dk_\rho =$$
$$= \frac{1}{8\pi\varepsilon_{r2}\varepsilon_0}\int_{-\infty}^{\infty}\tilde{J}_2(k_\rho)\kappa_2 H_0^{(1)}(k_\rho\rho)k_\rho dk_\rho$$
(42)

Therefore, from eqs. (41) and (42) we obtain the following system of algebraic equations :

$$\begin{cases}\dfrac{i\omega p k_\rho e^{i\kappa_1 x_0}}{\kappa_1}+\tilde{J}_1(k_\rho) = -\tilde{J}_2(k_\rho),\\ -i\omega p k_\rho e^{i\kappa_1 x_0}+\tilde{J}_1(k_\rho)\kappa_1 = \dfrac{\varepsilon_{r1}}{\varepsilon_{r2}}\tilde{J}_2(k_\rho)\kappa_2.\end{cases}$$
(43)

The solution of system of equations (43) are the uknown Fourier components of surface current densities, as following :

$$\tilde{J}_1(k_\rho) = i\omega p k_\rho e^{i\kappa_1 x_0}\frac{\varepsilon_{r2}\kappa_1 - \varepsilon_{r1}\kappa_2}{\kappa_1(\varepsilon_{r2}\kappa_1 + \varepsilon_{r1}\kappa_2)}$$

$$\tilde{J}_2(k_\rho) = -i\omega p k_\rho e^{i\kappa_1 x_0}\frac{2\varepsilon_{r2}}{\varepsilon_{r2}\kappa_1 + \varepsilon_{r1}\kappa_2}$$
(44)

*D. Expressions for the reflected and transmitted EM fields in integral representations*

Substituting expressions of eqs. (44) for the uknown current densities (at the interface, in spectral domain) in eqs. (27_ - (30), we obtain the reflected and transmitted EM fields in integral representations, as following :

In the higher half-space (LOS field plus reflected field, x>0):

$$\underline{H}(\underline{r}) = \underline{H}^{LOS} - \frac{i\omega p\,\hat{e}_\alpha}{8\pi}\int_{-\infty}^{\infty}\frac{\varepsilon_{r2}\kappa_1 - \varepsilon_{r1}\kappa_2}{\kappa_1(\varepsilon_{r2}\kappa_1 + \varepsilon_{r1}\kappa_2)}k_\rho^2\cdot$$
$$\cdot H_0^{(1)}(k_\rho\rho)e^{i\kappa_1(x_0+x)}dk_\rho$$
(45)

$$\underline{E}(\underline{r}) = \underline{E}^{LOS}(\underline{r}) - \frac{ip}{8\pi\varepsilon_{r1}\varepsilon_0}\hat{e}_\rho\int_{-\infty}^{\infty}k_\rho^2\frac{\varepsilon_{r2}\kappa_1 - \varepsilon_{r1}\kappa_2}{(\varepsilon_{r2}\kappa_1 + \varepsilon_{r1}\kappa_2)}\cdot$$
$$\cdot e^{i\kappa_1(x+x_0)}H_0^{(1)}(k_\rho\rho)dk_\rho +$$
$$+\frac{ip}{8\pi\varepsilon_{r1}\varepsilon_0}\hat{e}_x\int_{-\infty}^{\infty}k_\rho^3\frac{\varepsilon_{r2}\kappa_1 - \varepsilon_{r1}\kappa_2}{\kappa_1(\varepsilon_{r2}\kappa_1 + \varepsilon_{r1}\kappa_2)}e^{i\kappa_1(x+x_0)}\cdot$$
$$\cdot H_0^{(1)}(k_\rho\rho)dk_\rho$$
(46)

In the lower half-space (transmitted fields, x<0) :

$$\underline{H}^T(\underline{r}) = -\frac{i\omega p}{4\pi}\hat{e}_\alpha\int_{-\infty}^{\infty}k_\rho^2\frac{\varepsilon_{r2}}{\varepsilon_{r2}\kappa_1 + \varepsilon_{r1}\kappa_2}e^{i(\kappa_1 x_0 - \kappa_2 x)}\cdot$$
$$\cdot H_0^{(1)}(k_\rho\rho)dk_\rho$$
(47)

$$\underline{E}^T(\underline{r}) = -\frac{ip}{4\pi\varepsilon_0}\int_{-\infty}^{\infty}(k_\rho\hat{e}_x - \kappa_2\hat{e}_\rho)\frac{k_\rho^2}{\varepsilon_{r2}\kappa_1 + \varepsilon_{r1}\kappa_2}e^{i(\kappa_1 x_0 - \kappa_2 x)}\cdot$$
$$\cdot H_0^{(1)}(k_\rho\rho)dk_\rho$$
(48)

III. ELECTROMAGNETIC (EM) FIELDS REFLECTED FROM INFINITE, FLAT AND LOSSY GROUND IN THE FAR FIELD REGION : ANALYTICAL CLOSED – FORM EXPRESSIONS OBTAINED THROUGH THE APPLICATION OF THE STATIONARY PHASE METHOD (SPM).

In order to calculate the EM field above lossy ground (i.e. for x>0), we write eqs. (45) – (46) in the following form :

$$\underline{E}_{x>0} = \underline{E}^{LOS} - \frac{ip}{8\pi\varepsilon_o\varepsilon_1}I_1\cdot\hat{e}_\rho - \frac{ip}{8\pi\varepsilon_o\varepsilon_1}I_2\cdot\hat{e}_x$$
(49)

$$\underline{H}_{x>0} = \underline{H}^{LOS} - \frac{i\omega p}{8\pi}I_3\cdot\hat{e}_\alpha$$
(50)

where

$$I_1 = \int_{k_\rho=-\infty}^{\infty}\frac{\varepsilon_2\kappa_1 - \varepsilon_1\kappa_2}{\varepsilon_2\kappa_1 + \varepsilon_1\kappa_2}\cdot k_\rho^2\cdot H_o^{(1)}(k_\rho\rho)\cdot e^{i\kappa_1(x+x_o)}dk_\rho$$
(51)

$$I_2 = \int_{k_\rho=-\infty}^{\infty} \frac{k_\rho(\varepsilon_2\kappa_1 - \varepsilon_1\kappa_2)}{\kappa_1(\varepsilon_2\kappa_1 + \varepsilon_1\kappa_2)} \cdot k_\rho^2 \cdot H_o^{(1)}(k_\rho\rho) \cdot e^{i\kappa_1(x+x_o)} dk_\rho \tag{52}$$

and

$$I_3 = \int_{k_\rho=-\infty}^{\infty} \frac{\varepsilon_2\kappa_1 - \varepsilon_1\kappa_2}{\kappa_1(\varepsilon_2\kappa_1 + \varepsilon_1\kappa_2)} \cdot k_\rho^2 \cdot H_o^{(1)}(k_\rho\rho) \cdot e^{i\kappa_1(x+x_o)} dk_\rho \tag{53}$$

Furthermore, in order to calculate integral $I_1$ (in an almost identical manner integrals $I_2$ and $I_3$ will be calculated, using SPM method [6], [8]-[10]), let us assume large argument approximation for the Hankel functions of eqs. (51) – (53), namely let us assume that

$$k_\rho \cdot \rho \gg 1 \tag{54}$$

for which case function $H_0^{(1)}(k_\rho\rho)$ becomes a highly oscillating function of $k_\rho$. Then, since Stationary Phase Method (SPM) is to be applied, we just replace $H_0^{(1)}(k_\rho\rho)$ in eq. (51) by its asymptotic large argument approximation :

$$H_0^{(1)}(k_\rho\rho) = \sqrt{\frac{-2i}{\pi k_\rho\rho}} \cdot e^{+ik_\rho\rho} \tag{55}$$

Then integral $I_1$ of eq. (51) takes the following form :

$$I_1 = \sqrt{\frac{-2i}{\pi}} \cdot \frac{1}{\sqrt{\rho}} \int_{k_\rho=-\infty}^{\infty} k_\rho^{3/2} \cdot \frac{\varepsilon_2\kappa_1 - \varepsilon_1\kappa_2}{\varepsilon_2\kappa_1 + \varepsilon_1\kappa_2} \cdot e^{i\kappa_1(x+x_o)} e^{+ik_\rho\rho} dk_\rho \tag{56}$$

Moreover, in order to apply SPM method, we define radial distance $\rho$ (see Fig. 1) as 'large parameter', and we also define :

- 'Phase function' :

$$f(k_\rho) = \frac{\kappa_1(x+x_o)}{\rho} + k_\rho \tag{57}$$

- 'Amplitude function' :

$$F(k_\rho) = k_\rho^{3/2} \cdot \frac{\varepsilon_2\kappa_1 - \varepsilon_1\kappa_2}{\varepsilon_2\kappa_1 + \varepsilon_1\kappa_2} \tag{58}$$

Next, according to SPM method ([6], [8] – [10]) the 'stationary point' is calculated from the relation :

$$f'(k_\rho) = \frac{df(k_\rho)}{dk_\rho} = 0 \tag{59}$$

which finally yields the following expression for the 'stationary point' (only one stationary point exists) :

$$k_{\rho s} = \frac{k_{01}\rho}{\left[(x+x_0)^2 + \rho^2\right]^{1/2}} = k_{01}\frac{1}{\left[1+\left(\frac{x+x_0}{\rho}\right)^2\right]^{1/2}} \tag{60}$$

Note here that for the air - lossy ground problem considered here $k_{\rho s}$ is real and positive, and $k_{\rho s} < k_{01}$. Also, we can easily see that

$$\lim_{\rho\to\infty} k_{\rho s} = \lim_{(x+x_o)\to 0} k_{\rho s} = k_{01} \tag{61}$$

Furthermore, according to SPM method ([6], [8] – [10]), we also have to calculate the second derivative of the phase function, which in our case is calculated, from eq. (57), as

$$f''(k_{\rho s}) = -\frac{(x+x_o)}{\rho} \cdot \frac{k_{01}^2}{\left(k_{01}^2 - k_{\rho s}^2\right)^{3/2}} \tag{62}$$

Note here that $f''(k_\rho)$ is always negative, that is :

$$\text{sgn}\left[f''(k_{\rho s})\right] = -1 \tag{63}$$

which relation is needed in the application of SPM method. Then, by actually applying SPM method ([6], [8] – [10]), from eq. (56) we find :

$$I_1 = iF(k_{\rho s})e^{i\rho f(k_{\rho s})} \cdot e^{i\frac{\pi}{4}\text{sgn}[f''(k_{\rho s})]} \sqrt{\frac{2\pi}{\rho|f''(k_{\rho s})|}} \cdot \sqrt{\frac{2}{\pi\rho}} \exp(i\pi/4) \tag{64}$$

or

$$I_1 = \frac{i2}{\rho} \frac{1}{|f''(k_{\rho s})|^{1/2}} F(k_{\rho s}) e^{i\rho f(k_{\rho s})} \tag{65}$$

Then, by using expressions (57) – (58) and (62), we finally end up with the expressions :

$$I_1 = \frac{i2}{k_{01}\rho^{1/2}} \frac{1}{(x+x_0)^{1/2}} \kappa_{1s}^{3/2} k_{\rho s}^{3/2} \frac{\varepsilon_2\kappa_{1s} - \varepsilon_1\kappa_{2s}}{\varepsilon_2\kappa_{1s} + \varepsilon_1\kappa_{2s}} e^{ik_{\rho s}\rho} e^{i\kappa_{1s}(x+x_0)} \tag{66}$$

$$I_2 = \frac{i2}{k_{01}\rho^{1/2}} \frac{1}{(x+x_0)^{1/2}} \kappa_{1s}^{1/2} k_{\rho s}^{5/2} \frac{\varepsilon_2\kappa_{1s} - \varepsilon_1\kappa_{2s}}{\varepsilon_2\kappa_{1s} + \varepsilon_1\kappa_{2s}} e^{ik_{\rho s}\rho} e^{i\kappa_{1s}(x+x_0)} \tag{67}$$

$$I_3 = \frac{i2}{k_{01}\rho^{1/2}} \frac{1}{(x+x_0)^{1/2}} \kappa_{1s}^{1/2} k_{\rho s}^{3/2} \frac{\varepsilon_2 \kappa_{1s} - \varepsilon_1 \kappa_{2s}}{\varepsilon_2 \kappa_{1s} + \varepsilon_1 \kappa_{2s}} e^{ik_{\rho s}\rho} e^{i\kappa_{1s}(x+x_0)}$$

(68)

where

$$\kappa_{1s} = \sqrt{k_{01}^2 - k_{\rho s}^2}$$

(69)

$$\kappa_{2s} = \sqrt{k_{02}^2 - k_{\rho s}^2}$$

(70)

Then our final closed-form analytical solution consists of eqs. (49)-(50) and (66)-(70), where $k_{\rho s}$ is given by eq. (60).

## IV. PHYSICAL INTERPRETATION OF THE DERIVED CLOSED – FORM ANALYTICAL EXPRESSIONS FOR THE REFLECTED EM FIELDS

Regarding the physical interpretation of the derived solutions for the received EM field above the infinite, flat and lossy ground (which are *novel*, to our knowledge), eqs. (49) - (50), (60) and (66) – (70), we make the following remarks :

1. From eqs. (60)-(61) and (66) – (70), we can easily realize that *as $\rho \to \infty$ or as $(x + x_0) \to 0$ , i.e. very far away from the radiating dipole or very near the ground (i.e. 'surface wave') the EM waves propagate only parallel to the ground with wavenumber $k_{01}$.*

*In the general case*, always in the *far field region* examined in Section III, namely for

$$\rho \gg \lambda$$

(71)

as it is easily seen from the phase factors of eqs. (66) – (68) and exp(-iωt) time dependence assumed throughout this paper, *the EM fields propagate with wavenumber $k_{\rho s}$ in direction parallel to the ground, eq. (60) ($k_{\rho s} < k_{01}$), and with wavenumber $\kappa_{1s}$ in upwards direction [see eq. (69)]*. Then, by taking into account the horizontal and vertical wavenumbers of propagation mentioned just above, the total wavenumber of propagation equals to $k_{01}$, as derived through the use of eq. (69) :

$$k_{\rho s}^2 + \kappa_{1s}^2 = k_{01}^2$$

(72)

which, of course, had to be expected. *The above results, i.e. the values of wavenumbers of propagation of EM waves in the horizontal and vertical directions represent one of the interesting results of this paper.*

2. As $\rho \to \infty$ [and once $(x + x_0)$ is kept finite] it can be easily found from eqs. (66) – (70) that $k_{\rho s} \to k_{01}$, $\kappa_{1s} \to 0$,

$$\kappa_{2s} \to \sqrt{k_{02}^2 - k_{01}^2}$$ ,

the fractions appearing in eqs. (66) – (68) take (-1) value at the limit $\rho \to \infty$, and integrals $I_1$, $I_2$ and $I_3$ , as well as the radiated fields of eqs. (49) and (50), take the limiting value of zero, as expected.

3. For $\rho$= const. and $(x + x_0) \to 0$ [*surface wave behavior]*, similarly with case 2 above we find that $k_{\rho s} \to k_{01}$ , $\kappa_{1s} \to 0$ , the fractions appearing in eqs. (66) – 68) take (-1) value at the limit examined here, and :

$$I_1 \approx \frac{\kappa_{1s}^{3/2}}{(x+x_0)^2} \to 0$$

(73)

$$I_2 \approx \frac{\kappa_{1s}^{1/2}}{(x+x_0)^2} \to \frac{k_{01}^{1/2}}{\rho^{1/2}}$$

(74)

$$I_3 \approx \frac{\kappa_{1s}^{1/2}}{(x+x_0)^2} \to \frac{k_{01}^{1/2}}{\rho^{1/2}}$$

(75)

Then, as a conclusion, *in the case $(x + x_0) \to 0$ examined here [surface wave behavior]*, it can be easily realized from eqs. (49) and (50) that $E_x$ *(vertical polarization of electric field)* and $H_a$ *(azimuthal component of magnetic field) dominate in this case.*

4. In the case $(x + x_0) \gg \rho$ [i.e. radiating dipole and observation point well above the ground, in which case the *'space wave' dominates*], it can be easily shown that the *fractions in eqs. (66) – (68) take, in this limit, the value of the*
well known 'Fresnel reflection coefficient' $(n_1-n_2)/(n_1+n_2)$, where n represents the refractive index, $n = \sqrt{\varepsilon_r}$ .

5. Finally, regarding *frequency dependence of reflected EM fields*, note that from eqs. (66) – (68) integrals $I_1$ and $I_2$ vary in proportion to $\omega^2$ , while integral $I_3$ varies in proportion to $\omega$ . Then, finally, from eqs. (21) – (24) and (49) – (50) all EM fields (line of sight and reflected EM fields) vary in proportion to angular frequency $\omega$ [$\omega = 2\pi f$ , where f is the frequency of radiation], where the Hertzian dipole strength $I \cdot (2h) = p \cdot \omega$ (in magnitude) has been considered here as given (constant).

## V. CONCLUSIONS – FUTURE RESEARCH

In this paper we have derived analytical closed-form solutions for the received electromagnetic (EM) field for the problem of radiation of vertical Hertzian (small) dipole antenna above infinite, flat and lossy ground. To our knowledge these expressions are *novel* in the literature, and they have been derived here from a formulation in the spectral domain [4,5]. Futhermore, very interesting remarks regarding the physical interpretation of the analytical expressions mentioned above are presented in this paper, including wavenumbers of propagation (in horizontal and vertical directions), surface wave behavior and formula for the Fresnel reflection coefficient in the problem examined here, as well as in the limiting case of 'space waves' (where the usual expression for the Fresnel reflection coefficient is obtained).

Related research in the near future by our research group will include : comparison with EM field values to be obtained using K.A. Norton's approximate solutions [2], derivation of corresponding EM field expression for the transmitted EM field (region x < 0), solution of the corresponding problem for horizontal radiating Hertzian dipole above flat and lossy ground, propagation in isotropic and anisotropic crystals with interface (at x=0) etc.

REFERENCES

[1] A. N. Sommerfeld, "Propagation of Waves in Wireless Telegraphy", *Ann. Phys.,* 28, pp. 665 – 736, March 1909; and 81, pp. 1135 – 1153, December 1926.
[2] K. A. Norton, "The Propagation of Radio Waves Ovewr the Surface of the Earth", *Proceedings of the IRE,* 24, pp. 1367 – 1387, 1936; and 25, pp. 1203 – 1236, 1937.
[3] T. K. Sarkar et. al., "Electromagnetic Macro Modeling of Propagation in Mobile Wireless Communication : Theory and Experiment", *IEEE Antennas and Propagation Magazine,* Vol. 54, No. 6, pp. 17 – 43, Dec. 2012.
[4] S. Sautbekov, "The Generalized Solutions of a System of Maxwell's Equations for the Uniaxial Anisotropic Media', Chapter 1 in book *Electromagnetic Waves Propagation in Complex Matter,* edited by A. A. Kishk, Croatia, pp. 3 – 24, June 2011.
[5] S. Sautbekov, R. Kasimkhanova and P. Frangos, "Modified solution of Sommerfeld's problem", *Communications, Electromagnetics and Medical Applications (CEMA'10) International Conference*, National Technical University of Athens (NTUA), Athens, Greece, 7-9/10/2010, pp. 5 – 8.
[6] C. A. Balanis, *Antenna Theory : Analysis and Design,* Appendix VIII : Method of Stationary Phase, pp. 922 – 927, J. Wiley and Sons Inc., New York, 1997.
[7] J. Fikioris, *Introduction to Antenna Theory and Propagation of Electromagnetic Waves,* Book in Greek, National Technical University of Athens (NTUA), Athens, Greece, 1982.
[8] H. Moshovitis, *Asymptotic methods and Hogh Frequency Techniques for the Calculation of Electromagnetic Scattering by Using the Modified Stationary Phase Method,* Doctoral Dissertation, in Greek, National Technical University of Athens (NTUA), Athens, Greece, December 2010.
[9] Ch. Moschovitis, K. Karakatselos, E. Papkelis, H. Anastassiu, I. Ouranos, A. Tzoulis and P. Frangos, 'High Frequency Analytical Model for Scattering of Electromagnetic Waves from a Perfect Electric Conductor Plate using an Enhanced Stationary Phase Method Approximation', *IEEE Trans. Antennas and Propagation*, Vol. 58, No. 1, pp. 233 – 238, January 2010.
[10] Ch. G. Moschovitis, H. T. Anastassiu, and P. V. Frangos, 'Scattering of electromagnetic waves from a rectangular plate using an Extended Stationary Phase Method based on Fresnel functions (SPM-F*)'*, Progress In Electromagnetics Research (PIER), Vol. PIER 107, pp. 63-99, August 2010.
[11] G. Arfken, *Mathematical Methods for Physists,* 3$^{rd}$ Edition, pp. 400 – 414, Academic Press Inc., Orlando, Florida, USA, 1985.